\documentclass[12pt]{iopart}

\usepackage[lmargin=2.5cm,rmargin=2.5cm,tmargin=3cm,bmargin=2cm]{geometry}
\usepackage{setspace} % For adjusting line spacing
\usepackage{hyperref}
\hypersetup{
    colorlinks=true,
    linkcolor=blue,
    filecolor=blue,      
    urlcolor=blue,
    citecolor=blue
}
\usepackage{graphicx}
\usepackage[labelfont=up,textfont=up]{caption}
\usepackage{subcaption}
\usepackage{hhline}
\usepackage{url}
\usepackage{footmisc}
\usepackage{xcolor}
\usepackage{etoolbox}
\usepackage{titleps}
\usepackage{cite}
\usepackage{tabularx} % For better control over table column widths
\usepackage{verbatim} % For comments
\usepackage[capitalize]{cleveref}
\crefformat{footnote}{#2\footnotemark[#1]#3}
\usepackage{xspace}

\newcommand{\w}{\omega_0}
\newcommand{\g}{\gamma/\w}
\newcommand{\SI}[2]{#1~\mathrm{#2}}
\newcommand{\Aeff}{A_\mathrm{eff}}
\newcommand{\Aavg}{A_\mathrm{avg}}%{\bar{A}}%{A}

\newcommand{\halfwidth}{0.49\columnwidth}%{0.49\textwidth}

\newcommand{\Alfven}{Alfv\'{e}n\xspace}

\newcommand{\sub}[1]{\sout{#1}}
\renewcommand{\sub}[1]{\unskip}

\newcommand{\ak}[1]{\textcolor{red}{#1}}
\renewcommand{\ak}[1]{#1\unskip}

\newcommand{\nfadd}[1]{\textcolor{red}{#1}}
\renewcommand{\nfadd}[1]{#1\unskip}
\newcommand{\nfsub}[1]{\sout{#1}}
\renewcommand{\nfsub}[1]{\unskip}

\begin{document}
    
    \title[]{Isotope effects and Alfv\'en eigenmode stability in JET H, D, T, DT, and He plasmas}

%%% Affiliations %%%

\newcommand{\iPSFC}{$^1$\xspace}
\newcommand{\iCCFE}{$^3$\xspace}
\newcommand{\iEPFL}{$^4$\xspace}
\newcommand{\iCEA}{$^2$\xspace}
\newcommand{\iIPP}{$^6$\xspace}
\newcommand{\iKTH}{$^{5}$\xspace}
\newcommand{\iJET}{*\xspace}

\newcommand{\PSFC}{\iPSFC Plasma Science and Fusion Center, Massachusetts Institute of Technology, Cambridge, MA, USA\xspace}

\newcommand{\EPFL}{\iEPFL Ecole Polytechnique F\'{e}d\'{e}rale de Lausanne (EPFL), Swiss Plasma Center (SPC), CH-1015 Lausanne, Switzerland} % CH-1015

\newcommand{\CCFE}{\iCCFE Culham Centre for Fusion Energy, Culham Science Centre, Abingdon, UK} % OX14 3DB

\newcommand{\CEA}{\iCEA CEA, IRFM, F-13108 Saint-Paul-lez-Durance, France}

\newcommand{\UCI}{\iUCI Department of Physics and Astronomy, University of California, Irvine, California 92697, USA}

\newcommand{\NIFS}{\iNIFS National Institute for Fusion Science, Toki 509-5292, Japan}

\newcommand{\Ukraine}{\iUkraine Institute of Plasma Physics, NSC KIPT, 310108 Kharkov, Ukraine}

\newcommand{\Belgium}{\iBelgium Laboratory for Plasma Physics, LPP-ERM/KMS, TEC Partner, 1000 Brussels, Belgium}

\newcommand{\Milan}{\iMilan Dipartimento di Fisica, Universit\'{a} di Milano-Bicocca, 20126 Milan, Italy}

\newcommand{\IPST}{\iIPST Institute for Plasma Science and Technology, National Research Council, 20125, Milan, Italy}

\newcommand{\ESPCI}{\iESPCI Ecole Sup\'{e}rieure de Physique et de Chimie Industrielles de la Ville de Paris, 75231 Paris Cedex 05, France}

\newcommand{\Lisbon}{\iLisbon Instituto de Plasmas e Fus\~{a}o Nuclear, Instituto Superior T\'{e}cnico, Univ. de Lisboa, Lisbon, Portugal}%Instituto de Plasmas e Fus\~{a}o Nuclear, IST, Universidade de Lisboa, Lisbon, Portugal}

\newcommand{\PPPL}{\iPPPL Princeton Plasma Physics Laboratory, Princeton, NJ, USA}

\newcommand{\Slovenia}{\iSlovenia Jo\v{z}ef Stefan Institute, Ljubljana, Slovenia}

\newcommand{\Durham}{\iDurham Durham University, Durham, UK}

\newcommand{\IPP}{\iIPP Max-Planck-Institut für Plasmaphysik, Boltzmannstr. 2, 85748 Garching, Germany}

\newcommand{\KTH}{\iKTH Division of Fusion Plasma Physics, KTH Royal Institute of Technology, Stockholm, Sweden}

\newcommand{\JET}{\iJET See the author list of ``Overview of T and D-T results in JET with ITER-like wall'' by C.F. Maggi \emph{et al.} to be published in \emph{Nuclear Fusion Special Issue: Overview and Summary Papers from the 29th Fusion Energy Conference} (London, UK, 16-21 October 2023)}

\author{R.A.~Tinguely\iPSFC\footnote{Author to whom correspondence should be addressed: rating@mit.edu}\footnote{Shared first authorship.}, 
    P.G.~Puglia\iCEA\footnotemark[1],
    S.~Dowson\iCCFE,
    M.~Porkolab\iPSFC,
    D.~Douai\iCEA,
    A.~Fasoli\iEPFL,
    L.~Frassinetti\iKTH,
    D.~King\iCCFE,
    P.~Schneider\iIPP,
    and JET~Contributors\iJET}
    
    \address{\PSFC \\
             \CEA \\
             \CCFE \\
             \EPFL \\
             \KTH \\
             \IPP \\
             \JET}

    \begin{abstract}
    While much about \Alfven eigenmode (AE) stability has been explored in previous and current tokamaks, open questions remain for future burning plasma experiments, especially \ak{regarding} exact stability threshold conditions and related isotope effects\nfadd{; the latter, of course, requiring good knowledge of the plasma ion composition.} In the JET tokamak, eight in-vessel antennas actively excite stable AEs, from which their frequencies, toroidal mode numbers, and net damping rates are assessed. \nfadd{The effective ion mass can also be inferred using measurements of the plasma density and magnetic geometry.} Thousands of AE stability measurements have been collected by the Alfv\'en Eigenmode Active Diagnostic in hundreds of JET plasmas during the recent Hydrogen, Deuterium, Tritium, DT, and {Helium\ak{-4}} campaigns. In this novel AE stability database, spanning all four \ak{main} ion species, damping is observed to decrease with increasing Hydrogenic mass, but increase for Helium, a trend consistent with radiative damping as the dominant damping mechanism. These data are important for confident predictions of AE stability in both non-nuclear (H/He) and nuclear (D/T) operations in future devices. In particular, if radiative damping plays \ak{a} significant role in overall stability, some AEs could be more easily destabilized in D/T plasmas than their H/He reference pulses, even before considering fast ion and alpha particle drive. Active MHD spectroscopy is \nfadd{also} employed on select HD, HT, and DT plasmas to infer the effective ion mass, thereby closing the loop on isotope analysis \nfadd{and demonstrating a complementary method to typical diagnosis of the isotope ratio}.
\end{abstract}
    

    \renewcommand{\footnoterule}{\hrule width \linewidth}

    \section{JET's \Alfven Eigenmode Active Diagnostic}\label{sec:AEAD}

    \nfadd{
    Due to periodic boundary conditions, \Alfven waves can exist as \Alfven eigenmodes (AEs) in tokamak plasmas. The \Alfven frequency $\omega = 2\pi f$ comes from the simple shear \Alfven wave dispersion relation,
        \begin{equation}
            \omega = v_A k_\parallel = \frac{v_A}{R_0} \left(  n - \frac{m}{q} \right), 
            \label{eq:omegaA0}
        \end{equation}
    where $k_\parallel$ is the wave vector parallel to the magnetic field; $n$ and $m$ are toroidal and poloidal mode numbers, respectively; $q$ is the safety factor; and $R_0$ is the major radius. From this, the ``gap'' frequencies (within the \Alfven continua) of Toroidicity and Ellipticity-induced AEs are approximated as $\omega_\mathrm{TAE} \approx v_A/(2 q R_0)$ and $\omega_\mathrm{EAE} \approx v_A/(q R_0)$, respectively. Here, the \Alfven speed $v_A$ is given by
        \begin{equation}
            v_A = \frac{B_0}{\sqrt{ \mu_0 \sum_i{n_i m_i } }},
            \label{eq:vA0}
        \end{equation}
    with $B_0$ the on-axis toroidal magnetic field strength; $\mu_0$ the vacuum magnetic permeability; and $m_i$ and $n_i$ the ion masses and densities, respectively. From \cref{eq:omegaA0,eq:vA0}, it is clear how plasma parameters, like isotopic composition, can be inferred from measurements of AE frequencies, mode structures, and locations. 
    }
    
    \nfadd{While most magnetic confinement fusion experiments easily observe \emph{unstable} AEs,} the \Alfven Eigenmode Active Diagnostic (AEAD) actively probes \emph{stable} AEs in the JET tokamak \cite{Fasoli1995,Panis2010,Puglia2016}. Two toroidal arrays of antennas (four each, eight in total) are positioned below the midplane on the outboard wall inside the vacuum vessel \ak{and on opposite sides of the machine}. Each 18-turn antenna can be powered with currents up to ${\sim}\SI{10}{A}$, and the phasing of the antennas can be independently tuned so that the resulting perturbation has a dominant toroidal mode number $\vert n \vert < 20$. Four sets of filters, with different frequency bands, allow the AEAD frequency to span the range $f = \SI{25-330}{kHz}$, with TAEs typically excited in the range $f \approx \SI{100-300}{kHz}$ in JET plasmas \nfadd{\cite{Dumont2018,Fitzgerald2022}}.

    As the antenna drive scans through the resonant frequency of a stable AE, the plasma responds like a resonant cavity or a driven, damped harmonic oscillator. Arrays of high frequency magnetic probes measure this response, from which the eigenfrequency $\omega_0 = 2\pi f_0$ \nfadd{and} mode number $n$ \nfadd{ of the AE can be inferred, along with the} net damping rate $\gamma < 0$. Many damping mechanisms can contribute to the total damping rate: continuum~\nfadd{\cite{Rosenbluth1992,Zonca1992}}, radiative~\nfadd{\cite{Mett1992}}, collisional~\nfadd{\cite{Gorelenkov1992}}, Landau damping~\nfadd{\cite{Betti1992}}, and more. The mode can also experience drive, like that from fast ions, which pushes the mode toward instability, i.e. $\gamma > 0$ \nfadd{\cite{Berk1992,Breizman1995}}.%
        \footnote{\label{fn:NBI}\nfadd{
            In JET, fast ions from Neutral Beam Injection can \emph{damp} AEs due to their relatively low speed compared to the \Alfven speed \cite{Dumont2018}. %Actually, in JET, fast ions from Neutral Beam Injection often \emph{damp} AEs due to their relatively low speed compared to the \Alfven speed \cite{Dumont2018}.
        }}
    Thus, the alpha drive of AEs could be assessed directly by comparing two measurements in D and DT plasmas, assuming all other relevant plasma parameters are similar. \ak{Alpha drive could also be calculated via simultaneous measurements of two stable AEs with the same absolute $\vert n \vert$-value but opposite in sign, although this has not yet been observed.}
    %
    %    \footnote{
    %        Alpha drive could also be calculated via similar measurements of stable AEs with positive and negative $n$, although this has not been easily achieved.
    %    }

    An unambiguous measurement, by the AEAD, of alpha drive of a stable AE has not yet been identified for the recent JET DT campaigns\nfadd{, although a mode marginally \emph{destabilized} by alphas was detected \cite{Fitzgerald2023}.} Nevertheless, this system has successfully resonated with thousands of \ak{stable} AEs in hundreds of JET plasma discharges across multiple recent campaigns and with many novel observations. In this work, database trends of AE stability for various \ak{main} ion species are explored in Section~\ref{sec:database}.
    \ak{Importantly, a strong experimental case is made for the dominance of radiative damping in the data set, as well as the necessity to include the ion charge along with ion mass in theoretical scalings.}
    Furthermore, in Section~\ref{sec:isotope}, the ion mixture itself is inferred from AE frequencies \nfadd{via \emph{active} MHD spectroscopy, complementing visible light spectroscopy}. A summary and outlook are provided in Section~\ref{sec:summary}.
    \section{Isotope effects on AE stability}\label{sec:database}
%\section{Database analyses across JET H, D, T, DT, and He plasmas}\label{sec:database}

    %In this section, we investigate isotope effects on AE stability via a database of stable AE damping rates, $-\gamma < 0$.

    From 2019 through 2022, several single-species campaigns were carried out on JET to study isotope effects and \ak{to prepare for and} complement the 2021 DT campaign \cite{Mailloux2022}.%
        \footnote{The more recent 2023 JET DT campaign is outside the scope of the present work.}
    Across these Hydrogen, Deuterium, Tritium, and \ak{Helium-4}
        %\footnote{Only Helium-4, no Helium-3.}
    campaigns, a database of AE stability measurements was assembled with thousands  of data points from hundreds of plasma \ak{discharges}.
    %\ak{(Throughout this paper, we will abbreviate Helium-4 as ``He''; some JET discharges use Helium-3, but as a minority species.)}
    To the authors' knowledge, this publication is the first to explore AE stability across all four species of \ak{main} ions, i.e. mass numbers $A = 1-4$. Previous studies have investigated a subset of species and AEs: for example, TAEs in H, D, and T plasmas \cite{Fasoli2000pla} as well as H, D, and He discharges \cite{Testa2012} in JET; Beta-induced AEs (BAEs) in H and D plasmas in DIII-D \cite{Heidbrink2021}; and various \emph{fast} ion species in multiple tokamaks.
    \ak{(Throughout \nfadd{the rest of} this paper, we will abbreviate Helium-4 as ``He''; we note that some JET discharges use Helium-3, but as a minority species.)}
    
    %\ak{Note that, throughout this paper, we will abbreviate Helium-4 as ``He''; some JET discharges use Helium-3, but as a minority species.}

    In this section, we investigate isotope effects on general AE stability via the database of stable AE damping rates, $\gamma < 0$. 
    \nfadd{
    Specifically, we consider the measured damping rate \emph{normalized} by the measured eigenfrequency, i.e. $\g$, primarily because it enables clearer connection and comparison with theory, which often computes this more ``generalizeable'' quantity, whether for a TAE, EAE, or other AE. However, one caveat in this analysis is that the Doppler shift due to plasma rotation is \emph{not} used to correct in the frequency measurements in the database. The main reason for this is that plasma rotation profiles are not available for every plasma discharge, and uncertainties in the AE location and mode structure ($m,n$) would introduce further uncertainty in the subtracted Doppler shift. That said, the AEAD often struggles to probe AEs (i)~in the plasma core and (ii)~during high NBI power, so we expect low rotation for most data points.
    }

    \subsection{Trends with effective ion mass}\label{sec:mass}

    Because these five JET campaigns did not necessarily explore the same plasma parameter space, several filters are applied to the data set in order to consider a common subset: plasma currents $I_P = \SI{1-3}{MA}$; on-axis toroidal magnetic field strengths $B_T = \SI{1-2.5}{T}$; and minimal external heating powers ($P< \SI{1}{MW}$) from Neutral Beam Injection (NBI) and Ion Cyclotron Resonance Heating (ICRH); the last constraint is meant to minimize any contribution from fast ion drive \nfadd{(or damping)} and thereby isolate the effects of \nfadd{intrinsic} damping. \nfadd{Low NBI power often means low toroidal rotation as well, minimizing the Doppler shift and hence discrepancy between the measured AE frequency and ``true'' eigenfrequency $f_0$.} Uncertainties in the normalized damping rate and resonant frequency\nfadd{, from transfer function fits,} are also restricted to $\vert\Delta(\g)\vert <1\%$ and $\vert\Delta f_0 \vert<\SI{1}{kHz}$, respectively.

\begin{figure}[h!]
        \centering
        \captionsetup{font=it}
        \begin{subfigure}[t]{0.49\textwidth}
            \centering
            \includegraphics[width=\textwidth]{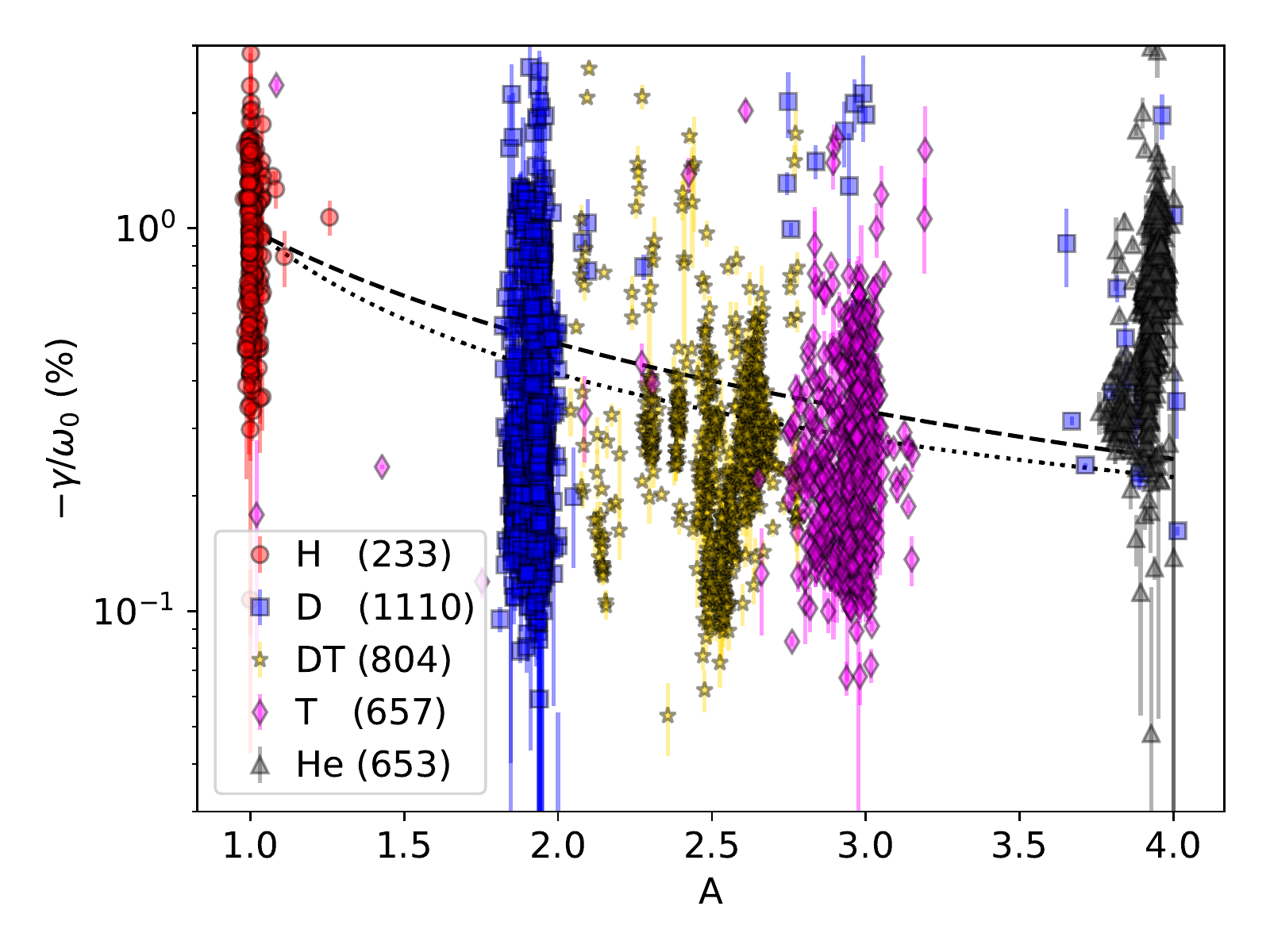}
            \caption{}
            \label{fig:scatter}
        \end{subfigure}
        \begin{subfigure}[t]{0.49\textwidth}
            \centering
            \includegraphics[width=\textwidth]{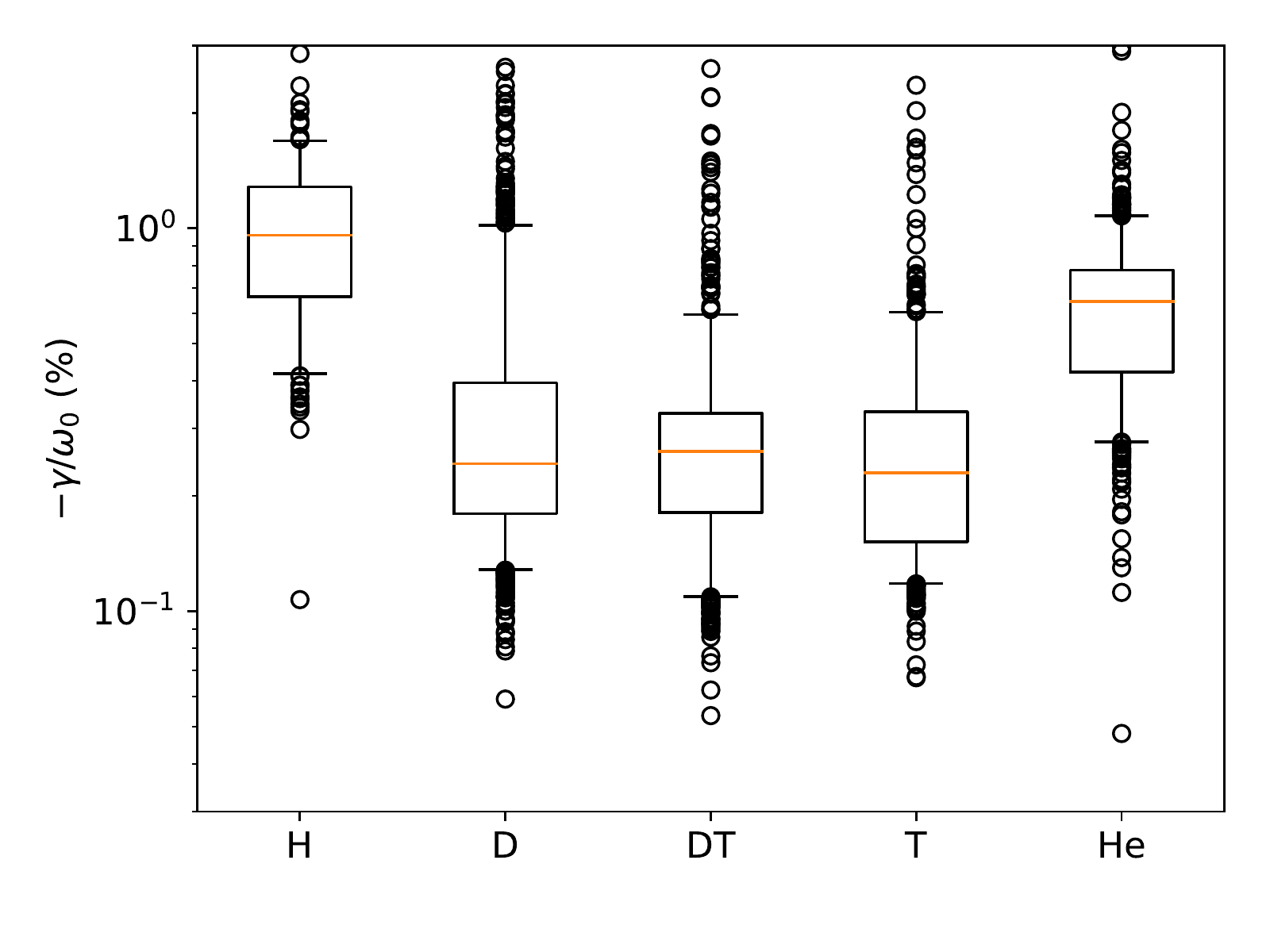}
            \caption{}
            \label{fig:box}
        \end{subfigure}
        \caption{(a)~Normalized damping rate $-\g$ vs average mass number $\Aavg$ for H (red circles), D (blue squares), DT (gold stars), T (magenta diamonds), and He (black triangles) plasmas with the number of data points in parentheses. Lines $-\g \propto 1/\Aavg$ \nfadd{\cite{Fasoli2000pla}} and $-\g \propto \exp(constant/\sqrt{\Aavg})$ \nfadd{\cite{Testa2012}} are dashed and dotted, respectively. (b)~Box plots for the same data: the yellow solid line indicates the median \nfadd{(50\%)}; box the first \nfadd{(25\%)} and third \nfadd{(75\%)} quartiles; \nfadd{bottom and top} ``whiskers'' the 5\% and 95\% quantiles, \nfadd{respectively;} and open circles the outliers.}
        \label{fig:database}
    \end{figure}

    Net damping rates, normalized to the resonant frequency, for all filtered data are shown in \cref{fig:scatter} as a function of the \emph{average} mass number $\Aavg = \sum_i A_i n_i / \sum_i n_i$ (with $n_i$ the ion density) assessed from edge spectroscopy. The data are well clustered around the expected mass numbers for each species and campaign. The slight skew toward lower $\Aavg$ values for D, T, and He plasmas \ak{could be} due to small fractions of H in some plasmas, e.g. as a minority species for ICRH.
    \ak{Impurities are also not taken into account for here, so the addition of their relatively high mass numbers could also help correct the leftward skew.}
    For the DT campaign, the D/T ratio was scanned more deliberately and can be seen as several distinct groupings. \ak{Despite the scatter,} a trend of decreasing damping rate with mass number is observed in the data for the Hydrogenic species ($\Aavg \approx 1-3$), approximately following an inverse relationship $-\g \propto 1/\Aavg$ (dashed in \cref{fig:scatter}). However, the trend fails for He plasmas which display higher damping rates than T data.

    This result is perhaps clearer in the distributions of data shown in \cref{fig:box}, where the median, 25\% and 75\% quartiles, and 5\% and 95\% quantiles are indicated for each campaign's data set. The median damping rate decreases with $\Aavg$ from H to D to T plasmas, but increases for He plasmas. 
    There are clear differences in the distributions of H vs D plasmas and T vs He plasmas.
    %\footnote{Note that the spread in each distribution could be due to variety of damping mechanisms: continuum, electron and ion Landau damping, and more; further investigations are left for future work.}
    The distributions of D, DT, and T data are much more alike; however, all quantiles are still lower for T than D data, with DT values falling somewhere in between. This confirms the inverse relationship of $-\g$ and $\Aavg$ for the Hydrogenic plasmas. Interestingly, the distributions for D and He plasmas are quite dissimilar, even though they have the same charge-to-mass ratio $Z/A$ and \nfadd{would} therefore \nfadd{be expected to} have similar \Alfven dynamics\nfadd{, i.e. \Alfven speeds and frequencies}.    %\ak{Apart from differences in plasma parameters, the spread in each distribution can also be due to variety of damping mechanisms: continuum, electron and ion Landau damping, and more; further investigations are left for future work.}  

    This trend of decreasing AE damping rate with main ion mass has been reported before. In \cite{Fasoli2000pla}, AE stability measurements across $\Aavg = 1-2.8$ were analyzed. Analytically, electron Landau damping was expected to scale as $-\g \propto \sqrt{\Aavg}$, disagreeing with experiment; however, gyrokinetic simulations matched the empirical finding of an inverse relationship, $-\g \propto 1/\Aavg$ \nfadd{(dashed line in \cref{fig:scatter})}, which was ultimately attributed to radiative damping via mode conversion to Kinetic \Alfven Waves.

    The study in \cite{Testa2012} complemented the previous analysis with data ranging $\Aavg = 1.7-3.95$. Similar decreasing trends were identified, although toroidal mode number discrimination was able to capture an increasing trend $-\g \propto \Aavg$. Importantly, the authors used an analytic expression for radiative damping with the form $-\g \propto \exp(G/\lambda)$~\cite{Connor1994}. Here, $G$ is a prefactor depending \ak{mainly} on the mode location in the \Alfven continuum, and $\lambda \propto q (\mathrm{d}q/\mathrm{d}r) \rho_i$ is a ``non-ideal'' parameter depending on the safety factor profile $q$, magnetic shear, and finite ion Larmor radius $\rho_i$. From this, the relationship $-\g \propto \exp(G'/\sqrt{\Aavg})$ was proposed in \cite{Testa2012}, which also matches the Hydrogenic data well in \cref{fig:scatter} (dotted line).

    Radiative damping can also explain the divergence of He data from the decreasing trend as the Larmor radius depends on both mass and charge: $ \rho_i \sim m_i v_{th,i}/qB \sim \sqrt{m_i T_i}/qB \propto \ak{\sqrt{A}/Z}$. Thus, a more appropriate expression for radiative damping can be written as $-\g \propto \exp(G' \, Z/\sqrt{\Aavg})$, with damping larger for He plasmas than D and T plasmas (since $2/\sqrt{4} = 1 > 1/\sqrt{2} > 1/\sqrt{3}$). However, this formula would imply that H and He data should have the \emph{same} damping (since $1/\sqrt{1} = 2/\sqrt{4}$), which is not reflected in \cref{fig:database}.
        %\ak{We note that the scatter in the data, or spread in each distribution, could be due to variety of damping mechanisms: continuum, electron and ion Landau damping, and more.}
    Simulations with MHD codes MISHKA \cite{mishka_mikhailovskii1997} and CASTOR \cite{cscas_huysmans2001} are underway to better evaluate dependencies on $n, Z, A,$ \nfadd{plasma rotation,} and the contributions of various damping mechanisms \ak{(e.g. continuum, electron and ion Landau damping, and more)}, which could explain this discrepancy \ak{and also the scatter in the data, or spread of each distribution}. \nfadd{Moreover, nonlinear and non-local interactions of multiple modes (e.g. core and edge-localized or multi-$n$) can add to damping \cite{Todo2012,Seo2021}.} Nevertheless, these raw damping rate data are important for future devices because, all else constant, AEs may be better \emph{stabilized} in pre-fusion-power operations with H/He plasmas and therefore more readily \emph{destabilized} in D/T campaigns. 

    \subsection{Trends with other plasma parameters}\label{sec:params}

    Dependencies of stable AE damping rates on other plasma parameters can also be explored for each campaign via the database. \Cref{tab:correlation} gives Pearson correlation coefficients $r_\mathrm{W}$, weighted by inverse variance, of the normalized damping rate with the edge safety factor $q_{95}$ and edge magnetic shear $s_{95}$, another formulation of the non-ideal parameter $\lambda' = q_{95} s_{95} \sqrt{T_\mathrm{e0}}/B_0$ \cite{Heidbrink2008} (with  $T_\mathrm{e0}$ and $B_0$ the on-axis electron temperature and toroidal magnetic field strength, respectively), and external heating powers from NBI and ICRH, across the different campaigns. 

    \nfadd{
    These parameters are selected for the following reasons: As seen in \cref{eq:omegaA0}, the shape of the \Alfven continua is strongly dependent on the $q$-profile; since the AEAD predominantly resonates with AEs localized near the plasma edge \cite{Tinguely2021}, $q_{95}$ (and to some extent $s_{95}$) can be a good indicator of continuum damping.\footnote{\nfadd{For examples of poor correlations with core-localized plasma values, e.g. central safety factor $q_0$, see Table~1 in \cite{Tinguely2022}.}} Furthermore, as discussed in \cref{sec:mass}, radiative damping is exponentially sensitive to $\lambda'$, with $\sqrt{T_\mathrm{e0}}/B_0$ serving as a proxy for the ion Larmor radius. Lastly, the $O(\SI{1}{MeV})$ fast ions accelerated by ICRH often drive AEs in JET, while the $O(\SI{100}{keV})$ fast ions from NBI can damp AEs \cite{Dumont2018} due to their relatively low speeds compared to $v_A$. 
    }% 
    Note that the data filter from \cref{sec:mass} is not used here in order to see trends including \emph{all} data within each campaign.

    \begin{table}[h!]
    \centering
    %\caption{\raggedright \MakeUppercase {{WEIGHTED LINEAR CORRELATION WITH NORMALIZED DAMPING RATE}}}
    %\caption{For different plasma ion species, weighted linear correlation coefficients ($r_\mathrm{W} \in [-1,1]$) of the normalized damping rate with various plasma parameters: edge safety factor, edge magnetic shear, non-ideal parameter of radiative damping theory, and NBI and ICRH powers. N/A is due to no NBI during the H campaign.}
    \caption{For different plasma ion species, weighted linear correlation coefficients ($r_\mathrm{W} \in [-1,1]$) of the normalized damping rate with various plasma parameters: edge safety factor, edge magnetic shear, non-ideal parameter of radiative damping theory, and NBI and ICRH powers. N/A indicates that no \ak{stable} AEs were detected with NBI during the H campaign.}
    \label{tab:correlation}
    \begin{tabular}{l r r r r r}%{l c c c c c}
        %\hline%{~~~}
        \hline%{-|-|-}
        $r_\mathrm{W}(-\g,\cdot)$ & $q_{95}$ & $s_{95}$ & $\lambda'$ & $P_\mathrm{NBI}$ & $P_\mathrm{ICRH}$ \\
        \hline%{-|-|-}
        H       & 0.79 & 0.65 & 0.76 & N/A        & $-$0.18 \\
        D \cite{Tinguely2022}    & 0.54 & 0.57 & 0.69 & 0.28     & $-$0.09 \\
        %T       & 0.71 & 0.70 & 0.74 & $-$0.11    & $-$0.04 \\
        \nfadd{DT}      & 0.46 & 0.61 & 0.65 & 0.10     & 0.14 \\
        \nfadd{T}       & 0.71 & 0.70 & 0.74 & $-$0.11    & $-$0.04 \\
        He      & 0.39 & $\sim$0.00 & 0.31 & $-$0.20    & 0.07 \\
        \hline%{-|-|-}
    \end{tabular}
\end{table}

    Strong positive correlations ($r_\mathrm{W}>0.5$) are observed with $q_{95}$, $s_{95}$, and $\lambda'$ for the Hydrogenic plasmas. This indicates that continuum and especially radiative damping are major contributors to AE stability for much of the data collected in these campaigns. In contrast, the He campaign shows a moderate correlation with $q_{95}$ and $\lambda'$, but essentially no correlation with $s_{95}$. 
        %\footnote{The range of $s_{95}$ data is narrower for He compared to the other campaigns, which could contribute to this vanishingly small correlation.}
    \ak{(We note here that the range of $s_{95}$ data is narrower for He compared to the other campaigns, which could contribute to this vanishingly small correlation.)}
    Thus, \nfadd{continuum and} radiative damping appear to be reduced \nfadd{for the He data, due to the smaller (positive) correlations}. This is consistent with the trend observed in \cref{fig:database} and could explain why the median damping rate is higher for H than He plasmas.
    
    %Thus, while continuum damping may still play a role, radiative damping appears to be reduced \nfadd{for the He data, due to the smaller (positive) correlations with $q_{95}$ and $\lambda'$}. This is consistent with the trend observed in \cref{fig:database} and could explain why the median damping rate is higher for H than He plasmas.

    Weak correlations ($\vert r_\mathrm{W} \vert<0.3$) are observed with NBI and ICRH powers,
        %\footnote{No NBI was used during the H campaign.}
    suggesting that they contribute minimally to the damping or drive of most measured stable AEs. This is likely due to the AEAD's improved coupling with the plasma edge \cite{Tinguely2021}, where modes have less interaction with primarily core-localized NBI or ICRH-accelerated fast ions. %However, the AEAD \emph{can} measure stable AEs with significant fast ion drive, as explored in the next section.

    \section{Isotope ratio measurements from stable AEs}\label{sec:isotope}

    MHD spectroscopy \cite{Goedbloed_1993,  Holties1997, SHARAPOV2001_127,Fasoli2002, Panis2010, Sharapov_Pellet_2018, Oliver_Pellet_2019} often utilizes \emph{passive} measurements of MHD instabilities to infer plasma properties.
    %; for example, locating a rational surface using magnetics measurements of neoclassical tearing modes. 
    The AEAD, on the other hand, is able to perform \emph{active} MHD spectroscopy, which has been demonstrated before on JET \cite{Fasoli2002,Testa2015}. In this work, our particular interest is measuring the \ak{main} ion composition, or isotope ratio in Hydrogenic plasmas, which will be of utmost importance in future DT fusion devices. The AEAD is well-suited for this as the \Alfven speed -- and hence the AE frequency -- depends on the \emph{effective} mass (described below) via $v_A \propto \Aeff^{-1/2}$. Moreover, destabilization by fast ions is not required to make this measurement, actually making the AEAD essential for this task when fast ion or alpha drive is insufficient to destabilize AEs.
        
    \begin{comment}
    To calculate $\Aeff$, we start with the simple shear \Alfven wave dispersion relation,
        \begin{equation}
            \omega = v_A k_\parallel = \frac{v_A}{R_0} \left(  n - \frac{m}{q} \right), 
            \label{eq:omegaA}
        \end{equation}
    where \nfadd{$\omega = 2\pi f$ is the mode frequency}, $k_\parallel$ is the wave vector parallel to the magnetic field; $n$ and $m$ are toroidal and poloidal mode numbers, respectively; $q$ is the safety factor; and $R_0$ is the major radius. From this, the TAE gap frequency is approximated as $\omega_\mathrm{TAE} \approx v_A/(2 q R_0)$. The \Alfven speed $v_A$ is given by
        \begin{equation}
            v_A = \frac{B_0}{\sqrt{ \mu_0 \sum_i{n_i m_i } }} = \frac{B_0}{\sqrt{ \mu_0 n_e m_{p} \Aeff }},
            \label{eq:vA}
        \end{equation}
    with $B_0$ the on-axis toroidal magnetic field; $\mu_0$ the vacuum magnetic permeability; $m_i$ and $n_i$ the ion masses and densities, respectively; $n_e$ the electron density; and $m_{p}$ the proton mass (Hydrogen). 
    \end{comment}
    %It is important to note that, with our definition of $\Aeff$ in \cref{sec:database}, the second equality in \cref{eq:vA} is only valid for \emph{Hydrogenic} plasmas, since the electron density $n_e = \sum_i n_i$ only when $Z_i = 1$. 
    %But electron density is routinely measured and is thus more convenient for our calculation.
    
    In this section, we assess the isotope ratio via stable AE frequencies in a few selected HD, HT, and DT discharges. Because the electron density $n_e$ is routinely measured in tokamaks, it is convenient to define an \emph{effective} mass number $\Aeff$ in relation to it. Assuming each \nfadd{ion} species $i$ is fully ionized, we have
        \begin{equation}\label{Aeff_eq}
            \Aeff  = \frac{\sum_i{A_i n_i}}{n_e} = \frac{\sum_i{A_i n_i}}{\sum_i{Z_i n_i}}.
        \end{equation}
    \nfadd{
    Therefore, we can rewrite the \Alfven speed from \cref{eq:vA0} as
        \begin{equation}
            v_A = \frac{B_0}{\sqrt{ \mu_0 n_e m_{p} \Aeff }},
            \label{eq:vA}
        \end{equation}
    %$v_A = B_0/\sqrt{ \mu_0 n_e m_{p} \Aeff }$
    where $m_{p}$ is the proton mass (Hydrogen).
    }
        
    It is important to note how the definitions of \emph{average} and \emph{effective} mass numbers are different: $\Aeff$ depends on $Z$ while $\Aavg$ (from \cref{sec:database}) does not; thus, for fixed $n_e$ (and $B_0$), all \ak{D and He4} plasmas \ak{(as well as any hypothetical mixed D-He4 plasmas)} have the same $\Aeff = 2$ and \Alfven speed, but \nfadd{a range of} $\Aavg = 2-4$. Conveniently, for Hydrogenic plasmas, $\Aeff = \Aavg$, but we maintain the distinction in the following analyses.

    %\red{But as electron density is routinely measured
    %It is convenient to define an effective mass (number) of the plasma $\Aeff$ in relation to it. Assuming a species $i$ is fully ionized, we have $n_{ei} = Z_i n_i$, $Z_i$ being the species atomic number, $\Aeff$ here should be defined as:}
        %\begin{equation}\label{Aeff_eq}
        %    \Aeff  = \frac{1}{n_e m_H} \sum_i{n_i m_i} = \frac{1}{n_e m_H} \sum_i{\frac{n_{ei} m_i}{Z_i} }
        %\end{equation}
    %\red{As can be seen from \cref{Aeff_eq}, the effective mass is dependent on the ratio between the atomic mass and charge of the different species; this means that, for example, a plasma composed of deuterium and helium-4 will have $A_{\mathrm{eff}} = 2$, when fully ionized, regardless of the proportion of each species.}

    %Note that  $n_e$ the electron density, and $m_H$ the hydrogen mass. %Commonly, 
    %The electron density $n_e$ is measured routinely on tokamak plasmas, so it is convenient to define an effective mass (number) of the plasma $A_{\mathrm{eff}}$ in relation to the electron density. Assuming a species $i$ is fully ionized, we have $n_{ei} = Z_i n_i$, $Z_i$ being the atomic number

    \nfadd{From \cref{eq:omegaA0,eq:vA},} we can relate the stable AE resonant frequency to the effective mass number via $f_0 \propto B_0/(q \sqrt{n_e \Aeff})$. Here, the challenge is identifying the proportionality constant which is dependent on the mode structure and location\nfadd{, although a simple multiplication of the prefactors, using $R_0 \approx \SI{3}{m}$ for JET, indicates that it should be of order $O(\SI{10^{12}}{kHz/T/m^{3/2}})$.} While we can often evaluate \nfadd{toroidal mode numbers} $n$ from experimental data, it is often more difficult to assess \nfadd{poloidal mode numbers} $m$; therefore, in the analyses below, we rely on (i)~plasmas being reproducible/similar and (ii)~the ability to choose a reference with known $\Aeff$.
    \ak{As in the previous section, impurities are not considered in the calculation, falling under assumption~(i) that similar plasmas have similar impurity contents; however, future work should investigate and include any isotope dependence of impurity accumulation.}

    \subsection{HD and HT isotope ratios}\label{sec:hdht}

\begin{figure}[h!]
        \centering
        \begin{subfigure}{\halfwidth}
            \includegraphics[width=\textwidth]{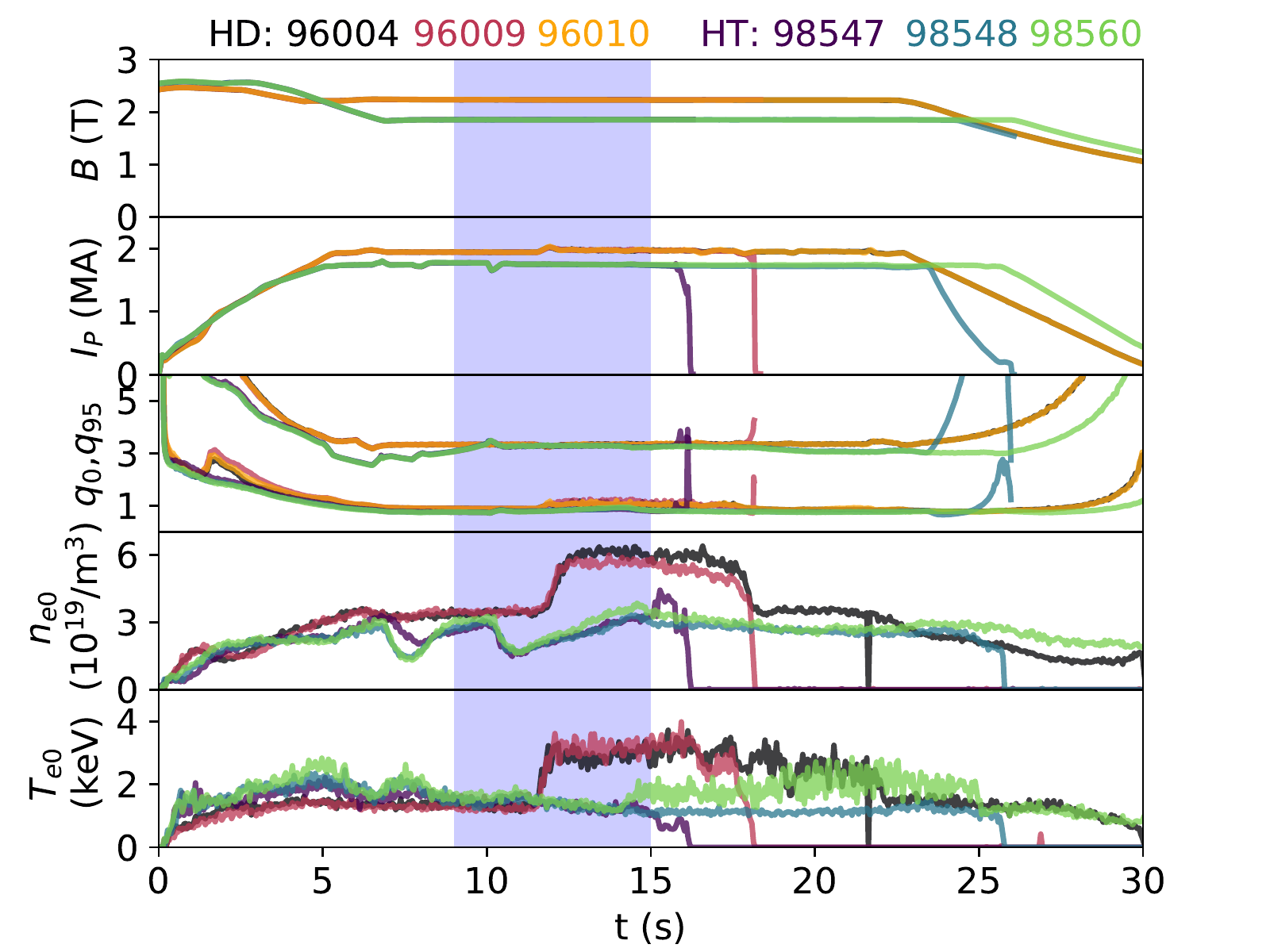}
            \caption{}
            \label{fig:params}
        \end{subfigure}
        \begin{subfigure}{\halfwidth}
            \includegraphics[width=\textwidth]{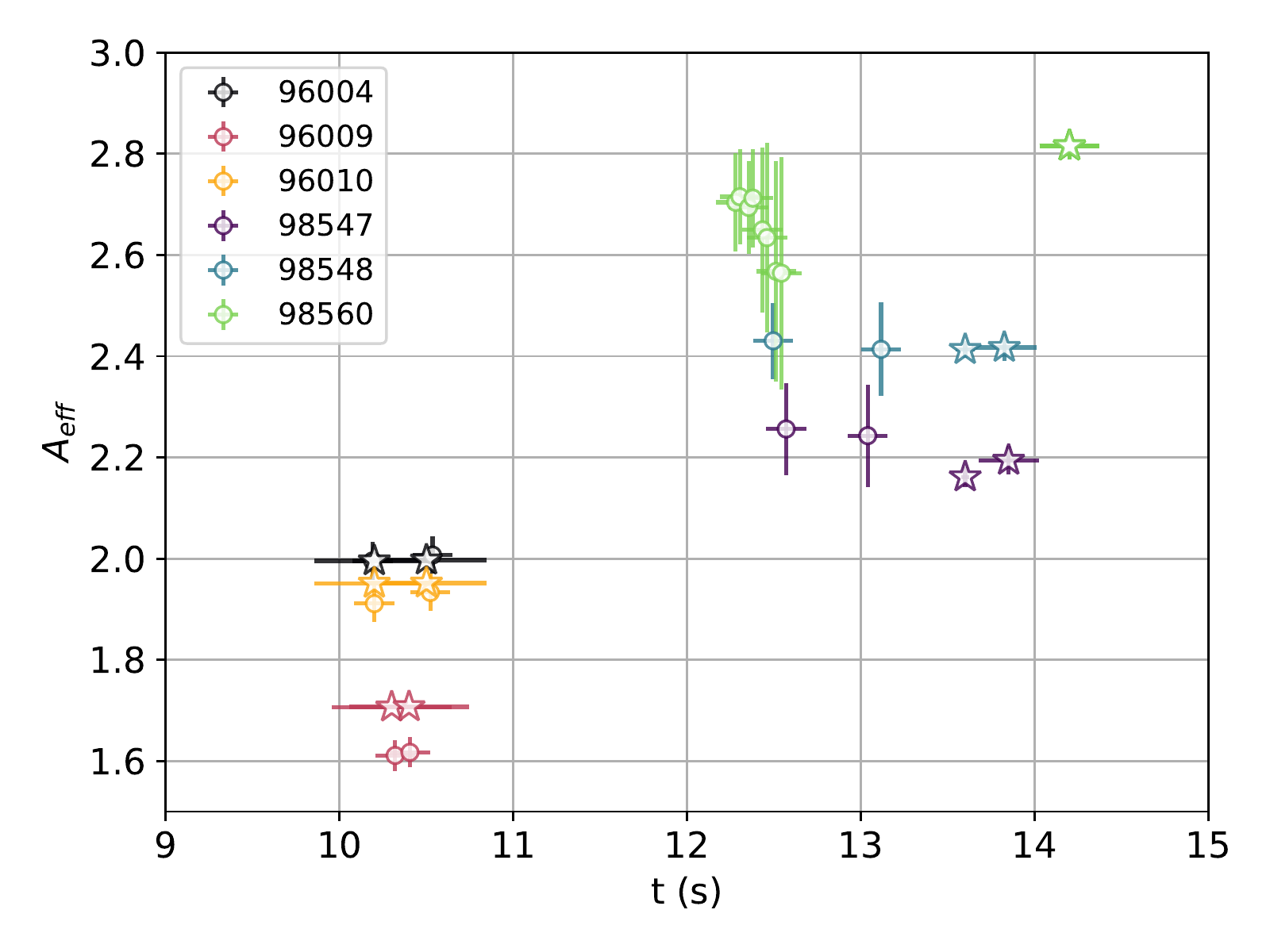}
            \caption{}
            \label{fig:HDT}
        \end{subfigure}
        \caption{(a)~Plasma parameters from six plasma discharges during which the isotope ratio is deduced from stable AEs (shaded region): on-axis magnetic field strength, plasma current, central and edge safety factors, and central electron density and temperature from Thomson scattering. (b)~Effective mass numbers estimated from AE frequencies (circles) and divertor spectroscopy (stars). \nfadd{HD discharges are JPN~96004-10; HT discharges are JPN~98547-60.} Note that \nfadd{Thomson} data is unavailable for JPN~96010.}
        \label{fig:params_and_HDT}
    \end{figure}

    Time traces of six plasma discharges, all part of a dedicated JET experiment on isotope ratio measurements in preparation for DT experiments, are shown in \cref{fig:params}. The magnetic geometries are well-reproduced in the pulses, evident in the similar toroidal magnetic field strengths, plasma currents, and central and edge safety factors. Two sets of three pulses were used to assess $\Aeff$ in HD and HT plasmas, respectively. The electron densities and temperatures are similar within those sets, especially during $t \approx \SI{9-15}{s}$ (shaded in \cref{fig:params}). \ak{The ICRH and NBI heating (not shown) are also comparable among the pulses in each set, but all start \emph{after} the \nfadd{AE} measurements of interest.}

    \Cref{fig:HDT} compares $\Aeff$ values calculated from divertor spectroscopy (stars) and from resonant AE frequencies in combination with global plasma parameters (circles): $B_0$ from magnetics, $q_{95}$ from EFIT \cite{Lao1985}, and \nfadd{$n_e$ from} line-integrated electron density. Here, JPN~96004 is selected as a pure (100\%) D reference plasma
    %, i.e. \nfadd{we impose} $\Aeff = 2$, 
    for \emph{both} the HD and HT plasmas\nfadd{, and again we assume that the AE is the same across these plasmas}. \nfadd{In other words, we choose $\Aeff = 2$ for JPN~96004 and utilize the above relation, $f_0 q \sqrt{n_e \Aeff}/B_0 \approx constant$, to calculate the $\Aeff$ values for the other discharges and times.} Note that the spectroscopic and stable AE measurements are not always simultaneous \ak{(unfortunately due to the different timings of the diagnostics)}. Even then, good agreement is observed between the two measurement methods, and most data agree either within error bars or at least $\sim$10\% uncertainty. This also provides further evidence that the AEAD primarily resonates with edge-localized AEs since the spectroscopic measurements also come from the edge plasma.
    
    %Because the spectroscopic measurement comes from the edge plasma and matches that from the AEAD, this is also consistent with the expectation and experience that many stable AEs excited by the AEAD also edge-localized. %Nevertheless, this nicely demonstrates that the AEAD can make a complementary measurement of the isotope ratio mixed isotope plasmas.
    \subsection{DT isotope ratio}

\begin{figure}[h!]
        \centering
        \includegraphics[width=0.95\columnwidth]{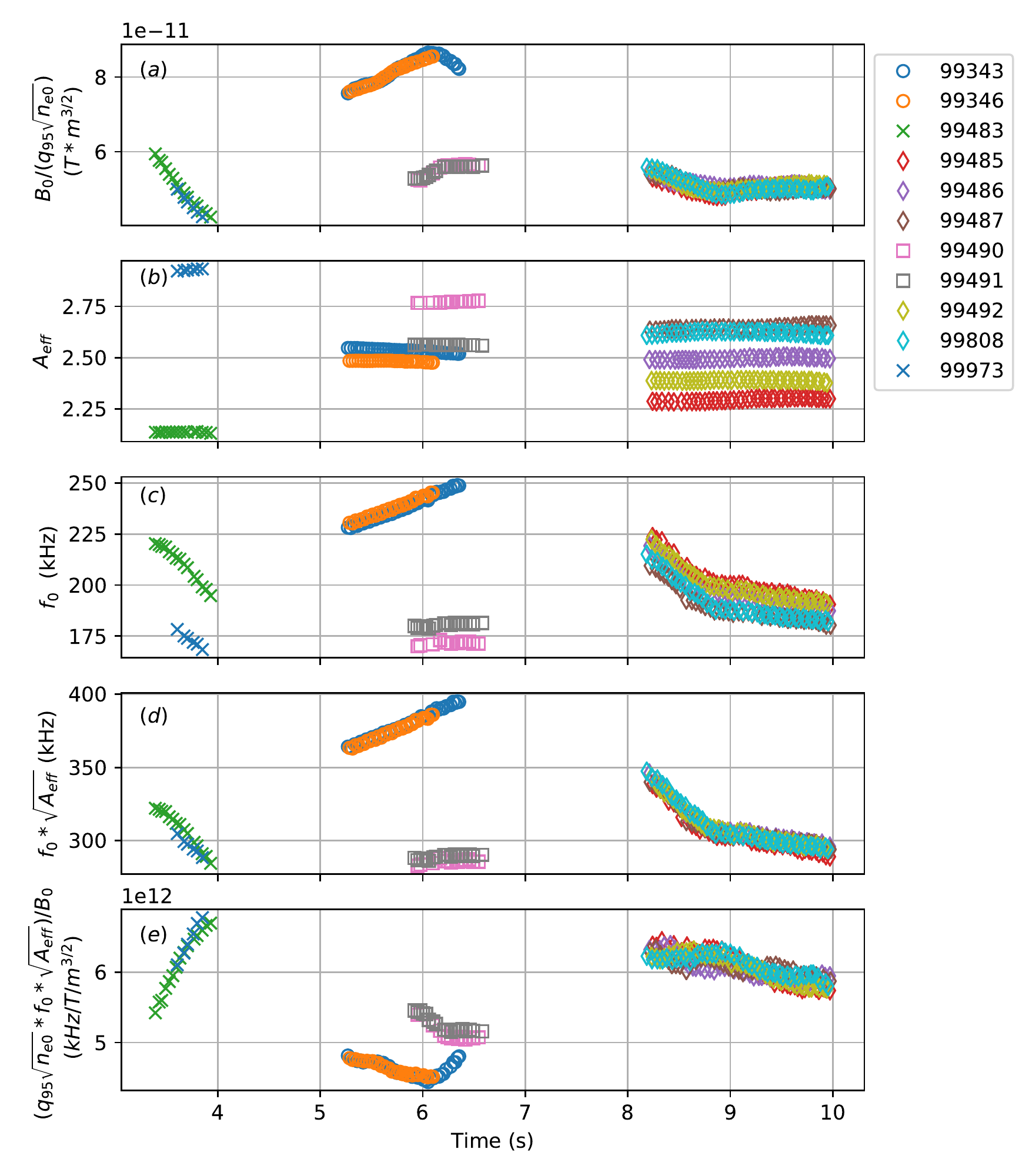}
        %\caption{Measurements showing the isotope effect on stable AE frequencies for similar JET pulses (grouped by symbol). From top to bottom: Relevant plasma parameters for the AE frequency dependency; effective masses from edge spectroscopy smoothed over $\SI{100}{ms}$; resonant frequencies measured by the AEAD; and multiplication of the square root of the effective mass by the stable AE frequency.}
        \caption{\nfadd{Measurements showing the isotope effect on stable AE frequencies for similar JET pulses (grouped by symbol). (a)~Relevant plasma parameters for the AE frequency dependency; (b)~effective masses from edge spectroscopy averaged over $\SI{100}{ms}$ intervals; (c)~resonant frequencies measured by the AEAD; (d)~multiplication of the square root of the effective mass by the stable AE frequency; and (e)~the ``constant'' of proportionality relating (a) and (d).}}
        \label{fig:Aeff_M21-12}
        \phantomsubcaption\label{fig:Aeff_M21-12a}
        \phantomsubcaption\label{fig:Aeff_M21-12b}
        \phantomsubcaption\label{fig:Aeff_M21-12c}
        \phantomsubcaption\label{fig:Aeff_M21-12d}
        \phantomsubcaption\label{fig:Aeff_M21-12e}
    \end{figure}

During the 2021 JET DT campaign, additional efforts were made to analyze the D/T isotope ratio using AEAD data. Similar to the previous section, this analysis focuses on comparing stable AE frequency measurements from alike pulses. However, here we do not choose a reference pulse, but instead show that spectroscopic measurements of $\Aeff$ are \emph{consistent} with those calculated from $f_0$. \Cref{fig:Aeff_M21-12} illustrates this, showing the effective mass of a series of pulses from the same experiment \cite{M21_12_Schneider_2023, M21_12_Frassinetti_2023}, each with varying D/T ratios. 

Within this experiment, distinct pulses have different ``clusters'' of stable AE resonances \ak{measured} during the ohmic phase, sometimes due to different plasma parameters but often simply due to the timing of the AEAD frequency scan. \ak{Here, the differences between pulse ``sets'' are different heating start times needed to optimize the $q$-profile and different plasma currents and magnetic field strengths relevant for the experiment: $(\SI{1.4}{MA}, \SI{1.7}{T})$ and $(\SI{2.0}{MA}, \SI{2.25}{T})$.} %
%The pulse differences were due to different heating starting time to optimize the q-profile and different plasma current and magnetic fields relevant for the experiment. 

\nfadd{\Cref{fig:Aeff_M21-12a}} shows the relevant plasma parameter combination\nfadd{, $B_0/(q_{95}\sqrt{n_{e0}})$,} for \Alfven waves at the times when stable AEs are excited by the AEAD. Note how these parameters are very similar \nfadd{(i.e. overlay each other)} for the reproduced pulse \ak{sets}. \nfadd{Here, we use $q_{95}$ since we expect the AEAD to resonate most strongly with edge AEs and because the edge safety factor ultimately shows a better correlation with the detected resonances than the central safety factor, $q_0$.}
\nfadd{\Cref{fig:Aeff_M21-12b}} shows the effective mass, again calculated from edge spectroscopy data and averaged over a $\SI{100}{ms}$ time window centered on each identified \nfadd{AE} resonance. 
The data span $\Aeff \approx 2.1 - 2.9$, a range which covers almost pure D to pure T plasmas. The \nfadd{measured} stable AE resonant frequencies are shown in \nfadd{\cref{fig:Aeff_M21-12c}}.

In \cref{fig:Aeff_M21-12d}, we simply multiply the square root of the effective mass by the AE resonant frequency. \nfadd{Because the plasma parameters in \cref{fig:Aeff_M21-12a} agree well among discharges, we also expect $f_0 \times \sqrt{\Aeff}$ to match (from \cref{eq:omegaA0,eq:vA}), which is observed. Importantly, the time evolution of data in both \cref{fig:Aeff_M21-12a,fig:Aeff_M21-12d} agree as well.}
The resulting data align nicely within 
%four distinct time windows: $t \approx \SI{3-4}{s}$, $\SI{5-6}{s}$, $\SI{6-7}{s}$, and $\SI{8-10}{s}$. 
three distinct time windows: $t \approx \SI{3-4}{s}$, $\SI{6-7}{s}$, and $\SI{8-10}{s}$. 
Within $t \approx \SI{3-4}{s}$, two traces with extremes in effective mass and time-varying frequencies converge to overlap in \nfadd{\cref{fig:Aeff_M21-12d}}. The later data highlight how even finer resolution in measuring $\Aeff$ is achievable for the plasmas spanning $\Aeff \approx 2.3-2.8$.

\nfadd{Lastly, in \cref{fig:Aeff_M21-12e}, we divide the data from \cref{fig:Aeff_M21-12d} by that in \cref{fig:Aeff_M21-12a}, which should give the proportionality constant dependent on the plasma profiles and AE mode structure. We find a value spanning ${\sim}\SI{4-7\times10^{12}}{kHz/T/m^{3/2}}$; this is higher than the aforementioned estimation of $10^{12}$. Yet this discrepancy could be due to the choice of plasma parameters used, e.g. edge safety factor $q_{95}$ and central electron density $n_{e0}$ (from Thomson scattering), since we do not know the precise mode location within the plasma (radius) and within the \Alfven continuum gap (frequency). Even though this ``constant'' evolves in time for each pulse set, it varies continuously and most importantly agrees among the pulses within each set. The relative variation is only $\sim$10\% for data after $t>\SI{5}{s}$; for $t<\SI{5}{s}$, the relative change is $\sim$30\%, probably due to the quickly changing plasma parameters. Comparing different pulse sets (or data ``clusters''), the variation in this proportionality factor could be due to different AEs excited by the AEAD and/or different mode locations due to the distinctly optimized $q$-profiles.}

It is important to acknowledge the limitations of the present analysis. As discussed, the AE frequency is influenced by various factors, and this approach to measuring the isotope ratio is most effective for similar plasma conditions. In fact,  due to the square root dependency, the AE frequency is twice as sensitive to a change $\Delta q/q$ compared to $\Delta \Aeff/\Aeff$ of the same magnitude. \ak{(The same goes for $B_0$ compared to $n_e$, but the toroidal magnetic field typically does not change much throughout a pulse.)}
    %\footnote{The same goes for $B_0$ compared to $n_e$, but the toroidal magnetic field typically does not change throughout a pulse.}
Thus, a steady magnetic configuration is needed.
Nevertheless, in future fusion power plants with a consistent plasma type and configuration, this method holds promise for future applications, offering a potential avenue for determining the isotope ratio in ongoing operations.

\section{Summary}\label{sec:summary}

    The recent single \ak{and mixed-}species campaigns at the JET tokamak, leading up to the 2021 DT campaign, have allowed the compilation of a comprehensive database of Alfv\'en Eigenmode (AE) stability across H, D, T, DT, and \ak{Helium-4} plasmas using the Alfv\'en Eigenmode Active Diagnostic (AEAD). Stable AE data from this recent He campaign are published here for the first time. This work may also be the first to analyze AE stability in plasmas with mass numbers spanning the full range $A = 1-4$. 
    
    As the mass number increases from H to D and DT to T campaigns, the distribution of normalized damping rates is found to shift toward lower values (see Fig.~\ref{fig:database}). This is consistent with previous isotope effect studies \cite{Fasoli2000pla,Testa2012} which attributed the decreasing trends, e.g. $-\g \propto 1/A$ or $\exp(1/\sqrt{A})$, to significant radiative damping. At first glance, the He data seem to contradict this conclusion as the distribution moves toward higher damping values; however, further inspection of radiative damping reveals a scaling $-\g \propto \exp(Z/\sqrt{A})$ with the introduction of charge $Z$ through the finite ion Larmor radius, a non-ideal/kinetic effect \cite{Connor1994}.

    An investigation of correlations within individual campaigns (see Table~\ref{tab:correlation}) further supports the dominance of radiative damping in the database. In particular, the damping rate is well correlated with the ``non-ideal parameter'' \cite{Heidbrink2008} associated with radiative damping. Strong correlations are found for Hydrogenic plasmas, whereas the moderate correlation for He plasmas could explain differences in the damping rate distributions of H and He plasmas, which otherwise have the same value of $Z/\sqrt{A}$. Regardless, these data indicate that radiative damping may play less of a role in AE stability for D/T plasmas than H/He plasmas; this could result in AEs being more easily destabilized in D/T discharges, even without alpha drive. Future tokamak operators must take this effect into account when moving from pre-nuclear to nuclear phases.

    Finally, the AEAD is shown to provide a reliable, complementary measurement of the isotope ratio, via stable AE frequencies, in Hydrogenic plasmas (see \cref{fig:params_and_HDT,fig:Aeff_M21-12}). This method is robust to plasma parameter changes of order 25\%/s and can even achieve a resolution $\Delta A \sim 0.1$ in similar plasmas. \ak{(It is not clear, however, if sufficient resolution could be achieved to assess low T-fractions during Tritium removal after a DT campaign \cite{Matveev2023}.)} Overall, these findings emphasize that active MHD spectroscopy is a viable tool to assess the fuel ion composition in future fusion devices, especially if AEs are \emph{not} destabilized in burning plasmas.

    %\section*{\hfill \textbf{ACKNOWLEDGEMENTS} \hfill}
\section*{Acknowledgements}

%Many thanks to the ICRH team and all JET Contributors \cite{Mailloux2022} for their experimental support. This work is supported by US DOE grants DE-SC0014264, DE-AC02-09CH11466, DE-SC0020412, DE-SC0020337, and Brazilian agency FAPESP Project 2011/50773-0. This research used resources of the National Energy Research Scientific Computing Center, a US DOE Office of Science User Facility operated under Contract No. DE-AC02-05CH11231 using NERSC award FES-ERCAP20598. This work has been carried out within the framework of the EUROfusion Consortium, funded by the European Union via the Euratom Research and Training Programme (Grant Agreement No 101052200 – EUROfusion). Views and opinions expressed are however those of the author(s) only and do not necessarily reflect those of the European Union or the European Commission. Neither the European Union nor the European Commission can be held responsible for them.

Many thanks to the JET ICRH team for their experimental support. This work is supported by US DOE grants DE-SC0014264, DE-AC02-09CH11466, DE-SC0020412, DE-SC0020337, and Brazilian agency FAPESP Project 2011/50773-0. This research used resources of the National Energy Research Scientific Computing Center, a US DOE Office of Science User Facility operated under Contract No. DE-AC02-05CH11231 using NERSC award FES-ERCAP20598. This work has been carried out within the framework of the EUROfusion Consortium, via the Euratom Research and Training Programme (Grant Agreement No 101052200 — EUROfusion) and funded by the Swiss State Secretariat for Education, Research and Innovation (SERI). Views and opinions expressed are however those of the author(s) only and do not necessarily reflect those of the European Union, the European Commission, or SERI. Neither the European Union nor the European Commission nor SERI can be held responsible for them.

     \section*{References}
        \bibliographystyle{unsrt}

    %\appendix
    %\input{density}

\end{document}